\documentclass[a4paper,11pt]{article}

\usepackage{jinstpub} 

\usepackage{lineno}

\usepackage{adjustbox}
\usepackage{comment}
\usepackage{xcolor}
\usepackage{multirow}

\usepackage{diagbox}

\usepackage{url}
\usepackage{hyperref}
\usepackage{cleveref}
\usepackage{upgreek}
\usepackage[]{siunitx}  
\usepackage[symbol]{footmisc}
\begin{document}

\title{GPU-based optical simulation of the DARWIN detector}

\author[a]{L.~Althueser}
\author[b,\ref*{banja}]{B.~Antunovi\'c}
\author[c]{E.~Aprile}
\author[d]{D.~Bajpai}
\author[e]{L.~Baudis}
\author[f]{D.~Baur}
\author[g]{A.~L.~Baxter}
\author[h]{L.~Bellagamba}
\author[i]{R.~Biondi}
\author[e]{Y.~Biondi}
\author[e]{A.~Bismark}
\author[f]{A.~Brown}
\author[j]{R.~Budnik}
\author[k]{A.~Chauvin}
\author[l]{A.~P.~Colijn}
\author[m]{J.~J.~Cuenca-Garc\'ia}
\author[i,n]{V.~D'Andrea}
\author[h]{P.~Di Gangi}
\author[f]{J.~Dierle}
\author[o]{S.~Diglio}
\author[k]{M.~Doerenkamp}
\author[m]{K.~Eitel}
\author[p]{S.~Farrell}
\author[i,n]{A.~D.~Ferella}
\author[i]{C.~Ferrari}
\author[d]{C.~Findley}
\author[f]{H.~Fischer}
\author[e]{M.~Galloway}
\author[e]{F.~Girard}
\author[f]{R.~Glade-Beucke}
\author[q]{L.~Grandi}
\author[r]{M.~Guida}
\author[k]{S.~Hansmann-Menzemer}
\author[r]{F.~J\"org}
\author[d,1]{L.~Jones\note[1]{Corresponding author, lajones18@crimson.ua.edu}}
\author[j]{P.~Kavrigin}
\author[s]{L.~M.~Krauss}
\author[m]{B.~von Krosigk}
\author[f]{F.~Kuger}
\author[j]{H.~Landsman}
\author[g]{R.~F.~Lang}
\author[g]{S.~Li}
\author[p]{S.~Liang}
\author[r]{M.~Lindner}
\author[o]{J.~Loizeau}
\author[t]{F.~Lombardi}
\author[r]{T.~{Marrod\'an Undagoitia}}
\author[o]{J.~Masbou}
\author[u]{E.~Masson}
\author[v,\ref*{coimbrapoly}]{J.~Matias-Lopes}
\author[b]{S.~Milutinovic}
\author[v]{C.~M.~B.~Monteiro}
\author[c]{M.~Murra}
\author[w]{K.~Ni}
\author[t]{U.~Oberlack}
\author[d,2]{I.~Ostrovskiy\note[2]{Corresponding author, iostrovskiy@ua.edu}}
\author[b]{M.~Pandurovic}
\author[e]{R.~Peres}
\author[g]{J.~Qin}
\author[f]{M.~Rajado Silva}
\author[e]{D.~Ram\'irez~Garc\'ia}
\author[e]{P.~Sanchez-Lucas}
\author[v]{J.~M.~F.~dos Santos}
\author[f]{M.~Schumann}
\author[h]{M.~Selvi}
\author[h]{F.~Semeria}
\author[r]{H.~Simgen}
\author[m]{M.~Steidl}
\author[x]{P.-L.~Tan}
\author[k,3]{A.~Terliuk\note[3]{Corresponding author, terliuk@physi.uni-heidelberg.de}}
\author[e]{K.~Thieme}
\author[y,\ref*{SISSA}]{R.~Trotta}
\author[p]{C.~D.~Tunnell}
\author[f]{F.~T\"onnies}
\author[m]{K.~Valerius}
\author[m]{S.~Vetter}
\author[e]{G.~Volta}
\author[d]{W.~Wang}
\author[e]{C.~Wittweg}
\author[o]{Y.~Xing}

\affiliation[a]{Institut f\"ur Kernphysik, Westf\"alische Wilhelms-Universit\"at M\"unster, 48149 M\"unster, Germany}
\affiliation[b]{Vinca Institute of Nuclear Science, University of Belgrade, Mihajla Petrovica Alasa 12-14. Belgrade, Serbia}
\affiliation[c]{Physics Department, Columbia University, New York, NY 10027, USA}
\affiliation[d]{Department of Physics and Astronomy, University of Alabama, Tuscaloosa, AL 35487, USA}
\affiliation[e]{Physik-Institut, University of Zurich, 8057  Zurich, Switzerland}
\affiliation[f]{Physikalisches Institut, Universit\"at Freiburg, 79104 Freiburg, Germany}
\affiliation[g]{Department of Physics and Astronomy, Purdue University, West Lafayette, IN 47907, USA}
\affiliation[h]{Department of Physics and Astronomy, University of Bologna and INFN-Bologna, 40126 Bologna, Italy}
\affiliation[i]{INFN-Laboratori Nazionali del Gran Sasso and Gran Sasso Science Institute, 67100 L'Aquila, Italy}
\affiliation[j]{Department of Particle Physics and Astrophysics, Weizmann Institute of Science, Rehovot 7610001, Israel}
\affiliation[k]{Physikalisches Institut, Ruprecht-Karls-Universit\"{a}t Heidelberg, 69120 Heidelberg, Germany}
\affiliation[l]{Nikhef and the University of Amsterdam, Science Park, 1098XG Amsterdam, Netherlands}
\affiliation[m]{Institute for Astroparticle Physics (IAP), Karlsruhe Institute of Technology (KIT), 76344 Eggenstein-Leopoldshafen, Germany}
\affiliation[n]{Department of Physics and Chemistry, University of L’Aquila, 67100 L’Aquila, Italy}
\affiliation[o]{SUBATECH, IMT Atlantique, CNRS/IN2P3, Universit\'e de Nantes, Nantes 44307, France}
\affiliation[p]{Department of Physics and Astronomy, Rice University, Houston, TX 77005, USA}
\affiliation[q]{Department of Physics \& Kavli Institute for Cosmological Physics, University of Chicago, Chicago, IL 60637, USA}
\affiliation[r]{Max-Planck-Institut f\"ur Kernphysik, 69117 Heidelberg, Germany}
\affiliation[s]{The Origins Project Foundation, Phoenix, AZ 85020, US}
\affiliation[t]{Institut f\"ur Physik \& Exzellenzcluster  PRISMA$^{+}$, Johannes Gutenberg-Universit\"at Mainz, 55099 Mainz, Germany}
\affiliation[u]{LPNHE, Universit\'{e} Pierre et Marie Curie, Universit\'{e} Paris Diderot, CNRS/IN2P3, Paris 75252, France}
\affiliation[v]{LIBPhys, Department of Physics, University of Coimbra, 3004-516 Coimbra, Portugal}
\affiliation[w]{Department of Physics, University of California, San Diego, CA 92093, USA}
\affiliation[x]{Oskar Klein Centre, Department of Physics, Stockholm University, AlbaNova, Stockholm SE-10691, Sweden}
\affiliation[y]{Department of Physics, Imperial Centre for Inference and Cosmology, Imperial College London, London SW7 2AZ, UK}
\collaboration{DARWIN collaboration}

\footnotetext[2]{Also at University of Banja Luka, Bosnia and Herzegovina\label{banja}}
\footnotetext[3]{Also at Coimbra Polytechnic - ISEC, Coimbra, Portugal\label{coimbrapoly}}
\footnotetext[4]{Also at SISSA, Data Science Excellence Department, Trieste, Italy\label{SISSA}}

\abstract{Understanding propagation of scintillation light is critical for maximizing the discovery potential of next-generation liquid xenon detectors that use dual-phase time projection chamber technology. This work describes a detailed optical simulation of the DARWIN detector implemented using Chroma, a GPU-based photon tracking framework. To evaluate the framework and to explore ways of maximizing efficiency and minimizing the time of light collection, we simulate several variations of the conventional detector design. Results of these selected studies are presented. More generally, we conclude that the approach used in this work allows one to investigate alternative designs faster and in more detail than using conventional Geant4 optical simulations, making it an attractive tool to guide the development of the ultimate liquid xenon observatory.}

\keywords{Dark matter; Double-beta decay; Chroma; Optical simulation; GPU}



\maketitle

\flushbottom

\section{Introduction}
\label{sec:intro}
Liquid xenon (LXe) time projection chambers (TPC) are a detector of choice for many experiments investigating dark matter~\cite{xenon1t_prl_2018_wimp,lux_prl_2017_complete}, neutrinoless double beta decay~\cite{exo_prl_2019}, and other physics channels~\cite{xenon_124xe_2019, x1t_excess,lux_axions}. In particular, dual-phase LXe TPC experiments have demonstrated tonne-level scalability~\cite{xenon1t_instrumental, LZ_instrumental,panda_x}, a keV-level energy threshold~\cite{xenon1t_s2only,lux_thr}, and sub-percent energy resolution at MeV energies~\cite{xenon1t_eres}. DARWIN is a planned third-generation (G3) detector~\cite{darwin2016} that will use $\sim$40 tonnes of natural xenon in a dual-phase TPC and aim at a tenfold increase in sensitivity to weakly interacting massive particles (WIMPs), compared to the current generation of experiments~\cite{XENONnT,LZ}. DARWIN's sensitivity to neutrinoless double-beta decay of $^{136}$Xe~\cite{darwin0nu} will be comparable to those of planned dedicated experiments~\cite{nexo}, due to having a similar mass of the isotope of interest (without the cost and risk of enrichment). A competitive measurement of solar neutrinos~\cite{darwin_solarNu} will also be possible.

The detection principle of a dual-phase LXe TPC is based on the measurement of the prompt scintillation light (S1) and the delayed proportional scintillation light emitted by the ionization electrons in the gas region (S2). An electric field is set up in the active region of the TPC between the negatively-biased cathode and the gate electrodes. The uniformity of the field is ensured by the field shaping rings (FSRs). The field drifts the electrons towards the liquid-gas boundary. A stronger electric field is set up between the gate and the positively-biased anode electrode, extracting the electrons into the gas region. The ratio of S2 to S1 signals (the charge-to-light ratio) depends on the interaction type and serves as a strong discriminator between nuclear recoil (NR) and electronic recoil (ER) events. 

Collecting as much of the scintillation light as possible is advantageous because it leads to better energy resolution, lower energy threshold, and improved ER/NR discrimination. Consequently, experiments strive to optimize their detectors for light collection, and developing a detailed Monte Carlo (MC) simulation of propagation and collection of scintillation light is critical for this step. The implementation of the optical simulation is commonly done using the Geant4 toolkit~\cite{geant4,baccarat,XENONnT}. It has been estimated that Geant4 photon tracking consumes $>$95\% of CPU time used in simulations of a current-generation dual-phase LXe TPC detector~\cite{opticks}. Recently, alternative MC frameworks are being considered for photon tracking by various groups~\cite{lixo,radial_tpc,single_phase,nexo,opticks} in the LXe community, offering a faster simulation and/or a more convenient geometry implementation. In this work, we describe a detailed optical simulation of the DARWIN detector implemented using one such framework, Chroma~\cite{chroma_snowmass,chroma_wp}. 

Chroma is a very fast optical simulation framework that was developed specifically for particle physics detectors. It runs on graphical processing units (GPUs) that have long been used for accelerating ray tracing by the computer graphics community~\cite{cs_1}. Chroma is claimed to be 50~\cite{chroma_wp} to 400~\cite{sorting} times faster than Geant4, which runs on central processing units (CPUs). Compared to a similar GPU-based framework~\cite{opticks,optiks_2019}, Chroma has an additional advantage of reducing the time to set up a complex simulation, which is particularly helpful during active detector development, when numerous alternative geometries need to be evaluated quickly. This is thanks to the ability to import detector geometries from computer-aided design (CAD) files. While Geant4 has a similar functionality, in practice it requires a process of converting objects from a CAD file using an external tool, followed by their import into Geant4 using the Geometry Description Markup Language (GDML) syntax. In contrast, Chroma directly imports a CAD file with a single internal method, making the process convenient and straightforward. Moreover, while the core of the framework's photon propagation code is written in CUDA C, the user interacts with Chroma using Python, taking advantage of the language's  flexibility and convenience. For an experiment that uses a Python-based modular framework in its full chain analysis, like DARWIN, integrating Chroma as one of the photon tracking libraries is a natural way to combine advantages of the Geant4's energy deposition simulation and Chroma's fast optical engine.

The above advantages of Chroma, which have already been utilized in several works~\cite{lixo, radial_tpc,single_phase,sorting,nexo,Theia}, motivated us to investigate the applicability of this framework for the development of the DARWIN detector. Given that Chroma is not yet as well-documented and supported as Geant4, this work first presents sanity checks and back-to-back comparisons in order to validate Chroma's performance. They are followed by several studies investigating light collection efficiency (LCE) and other metrics for the baseline configuration of the detector and for its potential modifications. While the primary goal of the studies is to evaluate the performance of Chroma and compare results to Geant4, they already provide valuable first insights into the possible design choices of the next-generation LXe TPC detector.

\section{Simulation of the DARWIN detector}
\subsection{Geometry in Chroma and Geant4}
\begin{figure}[htpb]
    \centering
    \includegraphics[width=0.95\textwidth]{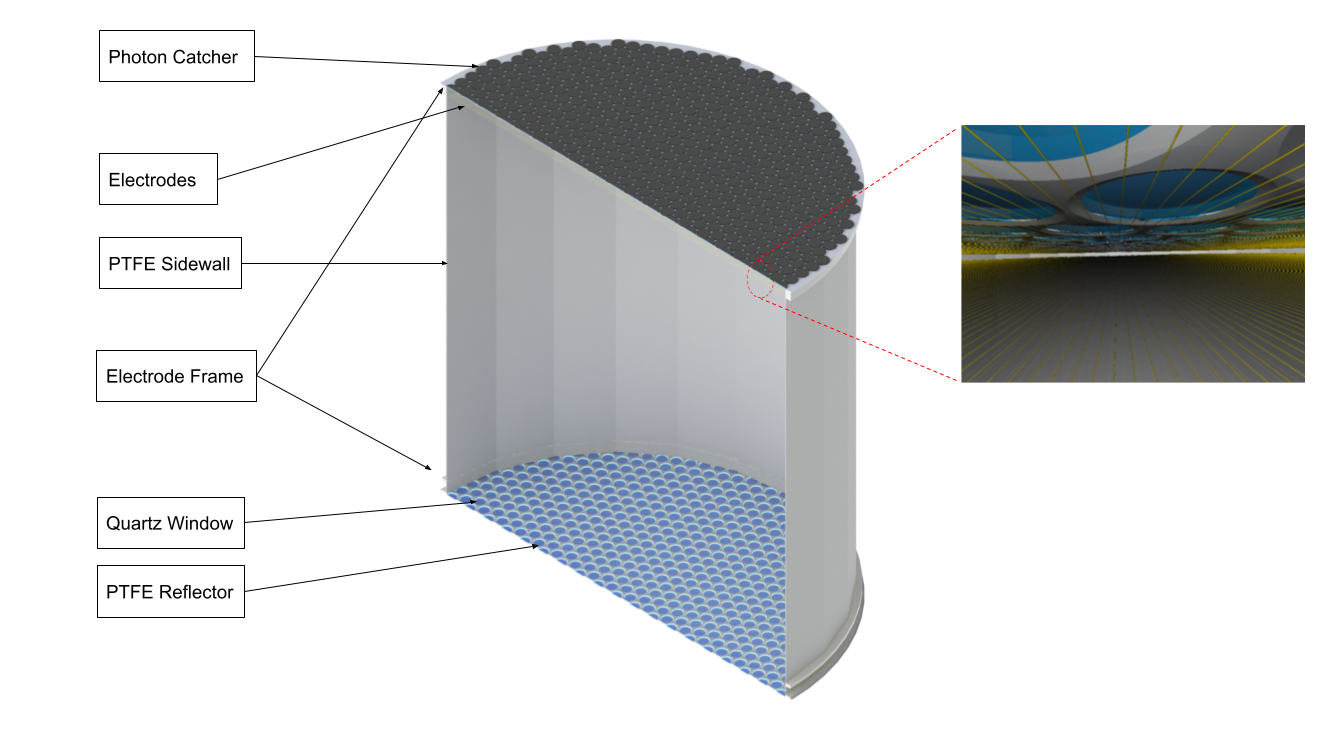}
    \caption{Cut-out view of the DARWIN detector, as imported into Chroma. Only components relevant for optical simulation are shown. The photocathodes of the bottom PMT array are seen as blue disks. The photon-absorbing caps of the top PMT array are seen in dark grey. The PTFE reflectors around the PMTs and on the sides, together with the electrode frames protruding from the sides, are shown in light grey. The diameter and height of the TPC are 2.6 m each. The Z axis direction is from the bottom to the top PMT arrays. The electrode wires cannot be seen in the main figure due to their small diameter (200 \si{\micro}m), so are shown in an insert.  The insert shows a Chroma rendering of some of the top PMTs (blue disks), the surrounding conical PTFE reflector (grey), and wires belonging to the three top electrodes (gold). The photocathodes are seen as dark blue disks behind the 5 mm thick quartz windows, shown in light blue.}
    \label{fig:cad}
\end{figure}

The baseline detector geometry is implemented in Chroma as a cylindrical dual-phase TPC with two arrays of 955 R11410-21 Hamamatsu photomultiplier tubes~\cite{pmt_europe,pmt_japan} (PMTs) each -- the top and bottom PMT arrays. The direction from the bottom to the top PMT arrays defines the Z axis of the detector's reference frame. The PMTs are closely packed in a hexagonal pattern to maximize light collection. The design of the PMTs (Figure~\ref{fig:cad}) includes only parts relevant for photon tracking and consists of a  64-\si{mm} diameter photocathode located 5 \si{mm} behind a quartz window and surrounded on all other sides by a photon-absorbing cap. The PMTs are surrounded by a polytetrafluoroethylene (PTFE) reflector that has a conical cross section to improve the light collection, as shown in the figure. PTFE reflectors are also placed on the sides of the TPC in the LXe and the gaseous xenon (GXe) regions. These sidewall reflectors are light-tight and have a shape of an extruded twenty-four-sided polygon inscribed inside a circle with a radius of 1300 mm. The height of the TPC, defined as the difference between the Z coordinates of the cathode and the gate, is 2598 mm ($\approx$2.6 m). The TPC contains three top and two bottom electrodes. In addition to the cathode, anode, and gate, which set up electric fields in the LXe and GXe regions, two screen electrodes shield the top and bottom PMT arrays. The electrodes are implemented as parallel stainless steel (SS) wires of 200 \si{\micro}m diameter and 5 (7.5) mm pitch for the top (bottom) case. The wires are fixed to SS frames that have a shape of an extruded twenty-four-sided polygon. The frames of the top electrodes are covered with PTFE, while the bottom frames are exposed to LXe. Mesh electrodes may be preferred in a large detector like DARWIN, and were also implemented as a cross-check. Only parts that could possibly be in the optical path of photons originated from the central volume of the TPC are imported into Chroma. In particular, the FSRs, which are placed immediately behind the sidewall reflectors, are not relevant for the baseline design. Figure~\ref{fig:cad} shows an overview of the baseline detector model, as imported in Chroma. Table~\ref{tab:z} shows the Z coordinates of some key components of the detector.  

\begin{table}[htpb]
\centering
\caption{Z coordinates of key components of the DARWIN detector as defined in the Chroma simulation. They are consistent with values used in the Geant4 simulation to 1-2 mm.}
\label{tab:z}
\begin{tabular}{|l|c|}
\hline
Component & Z (mm)\\
\hline
Photocathode (top PMT array) & 69 \\
Top Electrode 1 (Screen)  & 54 \\
Top Electrode 2 (Anode) & 12 \\
LXe/GXe boundary  & 2 \\
Top Electrode 3 (Gate)  & 0 \\
Bottom Electrode 1 (Cathode)  & -2598 \\
Bottom Electrode 2 (Screen)  & -2678 \\ 
Photocathode (bottom PMT array) & -2693\\
\hline
\end{tabular}
\end{table}

The default MC simulation pipeline in DARWIN utilizes the Geant4 toolkit (version 10.6.3.), which is the de facto standard in many areas of particle physics. While not imported from CAD files, as is done for Chroma, the geometry implementation follows the description above closely with only minor deviations. In particular, the sidewall PTFE reflector has a cylindrical shape in the Geant4 model, instead of the extruded twenty-four-sided polygon in Chroma. Two different photosensor implementations can be chosen during the detector construction: a simplified version with a single photosensitive disk spanning the whole area of a PMT array and the detailed version that models all Hamamatsu R11410-21 PMTs, with their internal structure, individually. The former option is used for non-critical simulations due to the speedup and allows one to easily study the impact of different sensor placement. The electrodes are implemented in Geant4 as thin solid disks, with a specified effective transparency. The transparency value was chosen based on estimates of shadowing of normally incident light by the electrode wires and is set to 95\% (97\%) for the top (bottom) electrodes. The dimensions and coordinates of the key components agree with those shown in Table~\ref{tab:z} to within 1-2~mm.

\subsection{Optical properties}
\label{sec:optical}
LXe scintillates in the vacuum ultraviolet region (VUV). The emission spectrum is modeled as a Gaussian with a mean of 174.8 nm and a full width at half maximum of 10.2 nm~\cite{LXe_wave}. The scintillation time profile is modeled by a two-component exponential decay function~\cite{tscint,Abe,Takiya,Hoge_1,Hoge_2,teresa}. 

Table~\ref{tab:optical} summarizes the optical properties at the mean wavelength, which are used for both the Geant4 and Chroma simulations. The refractive indices and extinction coefficients of relevant materials are labeled with $n$ and $k$, respectively. Only upper limits are established experimentally for the extinction coefficient of quartz~\cite{aop2007}, which is the material used in the PMTs' entrance windows. In the simulations we set this value to zero for simplicity, leading to no absorption of photons inside the quartz windows. The scintillation time constants and singlet fraction are labeled by $\tau_{1,3}$ and $f_{\mathrm{s}}$, respectively, and correspond to the assumed electric field value of 300 kV/cm.
\begin{table*}[htpb]
\centering
\caption{Main optical properties of materials used in the DARWIN simulation.}
\label{tab:optical}
\begin{adjustbox}{width=1\textwidth}
\begin{tabular}{|l|c|c|c|}
\hline
Parameter & Best Guess Value @175 nm & Spread  & Components\\
\hline
LXe refraction index, $n_{\mathrm{LXe}}$  & 1.69~\cite{nlxe_169} & 1.57~\cite{nlxe_157} -- 1.72~\cite{nim2002} & \multirow{9}*{LXe} \\ \cline{1-3}
Scint. mean wavelength, nm & 174.8~\cite{LXe_wave} &  & \\ \cline{1-3}
Scint. FWHM, nm & 10.2~\cite{LXe_wave} & & \\ \cline{1-3}
Scint. singlet time constant, $\tau_1$, ns & 3.5  & 2.4~\cite{tscint} -- 4.3~\cite{Abe} & \\ \cline{1-3}
Scint. triplet time constant, $\tau_3$, ns & 24  & 23~\cite{tscint} -- 37~\cite{Takiya} & \\ \cline{1-3}
Scint. NR singlet fraction, $f^{\mathrm{NR}}_{\mathrm{s}}$ & 0.66 & 0.25~\cite{Abe} -- 0.69~\cite{tscint}  & \\ \cline{1-3}
Scint. ER singlet fraction, $f^{\mathrm{ER}}_{\mathrm{s}}$ & 0.22 & 0.12~\cite{Hoge_2} -- 0.25~\cite{tscint}  & \\ \cline{1-3}
LXe absorption length, L$_{\mathrm{abs}}$, m  & 50~\cite{XENONnT} & 20~\cite{nexo_sens} -- 100~\cite{LZ} & \\ \cline{1-3}
LXe scattering length, L$_{\mathrm{Rayleigh}}$, cm  & 36~\cite{ray_2} & 20~\cite{nim2002} -- 50~\cite{nim2002} &  \\ \hline
GXe absorption length, L$_{\mathrm{abs}}$, m  & 500 &  & \multirow{2}*{GXe} \\ \cline{1-3}
GXe scattering length, L$_{\mathrm{Rayleigh}}$, m  & 200 &    & \\ \hline
PTFE reflectivity in LXe, $R_{\mathrm{total}}$, \%  & 99 & 93~\cite{LZ} -- 100 & PTFE (side, bottom)   \\ \hline
PTFE reflectivity in GXe, $R_{\mathrm{total}}$, \% & 80 & 75~\cite{LZ} -- 85~\cite{LZ} & PTFE (side, top)\\ \hline
Specular/diffuse reflectivity ratio for PTFE, $R_{\mathrm{spec}}$/$R_{\mathrm{diff}}$ & 0.05 & 0.05~\cite{coimbra} -- 0.35~\cite{coimbra} & PTFE  (side, top, bottom)\\ \hline
Quartz refraction index, $n_{\mathrm{quartz}}$  & 1.60 & 1.55~\cite{aop2007} -- 1.69~\cite{aop2007} & \multirow{2}*{PMT window} \\ \cline{1-3}
Quartz extinction coefficient, $k_{\mathrm{quartz}}$ & $<10^{-4}$ & $<10^{-2}$~\cite{aop2007} -- $<10^{-6}$~\cite{aop2007} & \\ \hline
SS reflectivity, $R_{\mathrm{total}}$ \% & 30~\cite{r_ss} &  & \multirow{2}*{Electrodes and frames} \\ \cline{1-3}
Specular/diffuse reflectivity ratio for SS, $R_{\mathrm{spec}}$/$R_{\mathrm{diff}}$ & 2 &  & \\ \hline
\end{tabular}
\end{adjustbox}
\end{table*}
Reflectivity is denoted by $R$, separately for diffuse and specular components. The table also specifies which components are assigned the corresponding properties. The best guess spread in the property values is also shown when available. Insufficient reliable information is available for many of the components, which motivates additional measurements in spite of the difficulty of conducting studies in VUV and in the LXe environment. Reflective properties of some surfaces may also vary substantially depending on the surface finish and the state of oxidation~\cite{silva2007}. 

Surface reflectivity in Chroma can either be explicitly specified (separately for diffuse and specular components), or calculated using Fresnel equations from the specified complex index of refraction and surface thickness. In the former case, the reflectivity is assumed to be independent of the angle of incidence. It is known that reflectivity in LXe is not always aligned with the expectation based on the refractive index value measured in vacuum~\cite{coimbra,coimbra2,guofu}. Oxidizing metal surfaces and wet PTFE are prime examples. Additionally, the composition of the surface layers is unknown for some types of photosensors used in this work, like Hamamatsu VUV4 silicon photomultipliers (SiPMs), making calculation of the expected reflectance impossible. Consequently, the only components whose reflectivity are calculated from the refractive indices in our simulation are quartz, LXe, and GXe. 


\section{Comparison of Chroma and Geant4}
\label{sec:comparison}
To investigate the reliability of Chroma simulations, checks and comparisons with Geant4 were performed, as described below. During these checks, a few issues with the Chroma code were uncovered, which accounted for a few percent discrepancy between the simulations. The issues and how they were fixed in our local installation of Chroma are described in detail in Appendix. This section describes the comparisons performed after the fixes have been implemented.

In the first two checks, we investigated refraction and angular-dependent reflection on the LXe/GXe boundary and on the boundaries between xenon and PMT entrance windows (quartz) by simulating 10$^5$-10$^6$ photons directed at these interfaces and counting how many of them get detected by the top and bottom PMT arrays. Absorption and scattering processes were turned off, the electrodes were removed, and the photon detection efficiency of the PMTs was set to 100\% for these checks. 

When directing photons up from under the liquid level of the detector, 6.5\% of photons should get reflected back from the LXe/GXe interface. Of the 93.5\% photons that pass into the gas phase, 4.6\% should reflect from the GXe/quartz and pass through GXe/LXe on the way down. Overall, 11.2\% (88.8\%) of the 10$^5$ generated photons are predicted to be detected by PMTs in the bottom (top) arrays. Both Chroma and Geant4 agree with the prediction within statistical uncertainties.

\begin{figure*}[htp]
    \centering
    \includegraphics[width=0.9\textwidth]{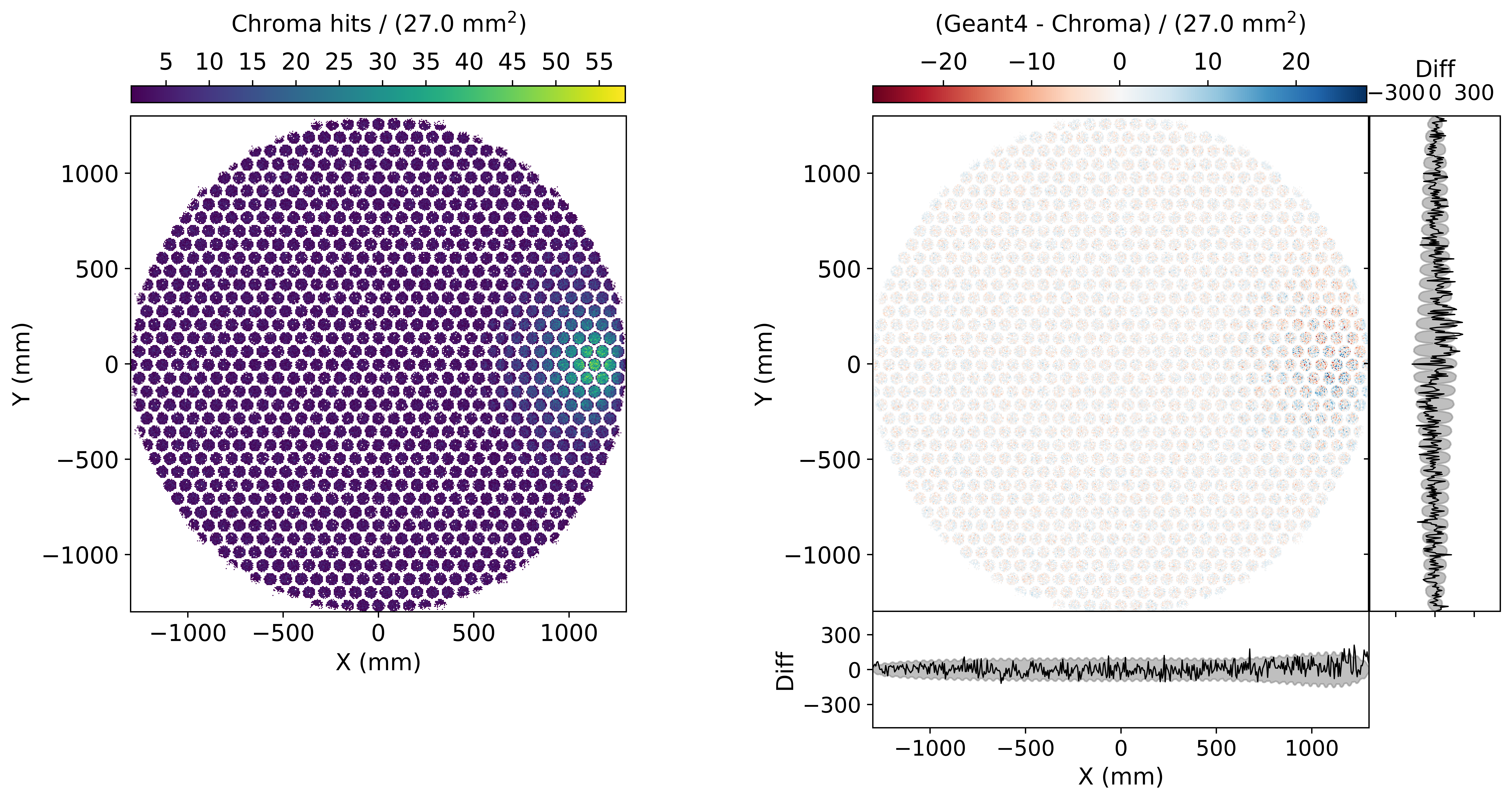}
    \caption{(Left) Distribution of positions of the 10$^6$ photons that were directed at a 75-degree angle of incidence on a bottom PMT, underwent total internal reflection on the LXe/quartz interface, and reflected from the PTFE sidewall before being detected by the bottom PMT array in Chroma. (Right) Difference between the Geant4 and Chroma distributions. Projections on the X and Y axis shown on the right and bottom panels. The grey bands show the regions around zero with the width equal to $\pm$2$\sigma$ statistical error. The differences are centered around zero indicating a good agreement between the two frameworks.}
    \label{fig:total_reflection}
\end{figure*}

When directing photons towards a bottom PMT at 75 degrees of incidence, one expects that 100\% of the 10$^6$ photons would get reflected off the PMT's quartz window. Both Chroma and Geant4 are confirmed to simulate the total internal reflection at the LXe/quartz interface. The photons then hit the PTFE sidewall at an expected Z coordinate and (mostly diffusely) reflected. Distribution of positions of photons that eventually got detected by the bottom PMT array are shown in Figure~\ref{fig:total_reflection}. The figure also shows the difference between the Geant4 and Chroma distributions, together with the projections on X and Y axes. For this test, a cylindrical PTFE sidewall was imported in Chroma to match the geometry in Geant4. The results of the two frameworks are in good agreement. 
A similar test also confirmed the total internal reflection at the LXe/GXe interface. The total internal reflection occurs at the expected critical angle of incidence for both frameworks.

The next check investigated LCE as a function of the Z coordinate. LCE is defined here as the number of photons hitting the photosensitive area divided by the total number of generated photons. Therefore, the definition of LCE includes potential reflections from the photosensors’ entrance windows. This definition does not include the photon detection efficiency, which for different photosensors varies from O(20\%)~\cite{vuv4} to O(35\%)~\cite{2in}. The absorption and scattering lengths were now set to their best guess values. 10$^5$ photons were generated isotropically at $\vec{r}$=(0,0,Z), where Z varied from -2648 (50 mm below the cathode position) to 52 mm (50 mm above the LXe/GXe interface position). Figure 4 shows the Z dependence of the LCE.
\begin{figure}[htp]
    \centering
    \includegraphics[width=0.65\textwidth]{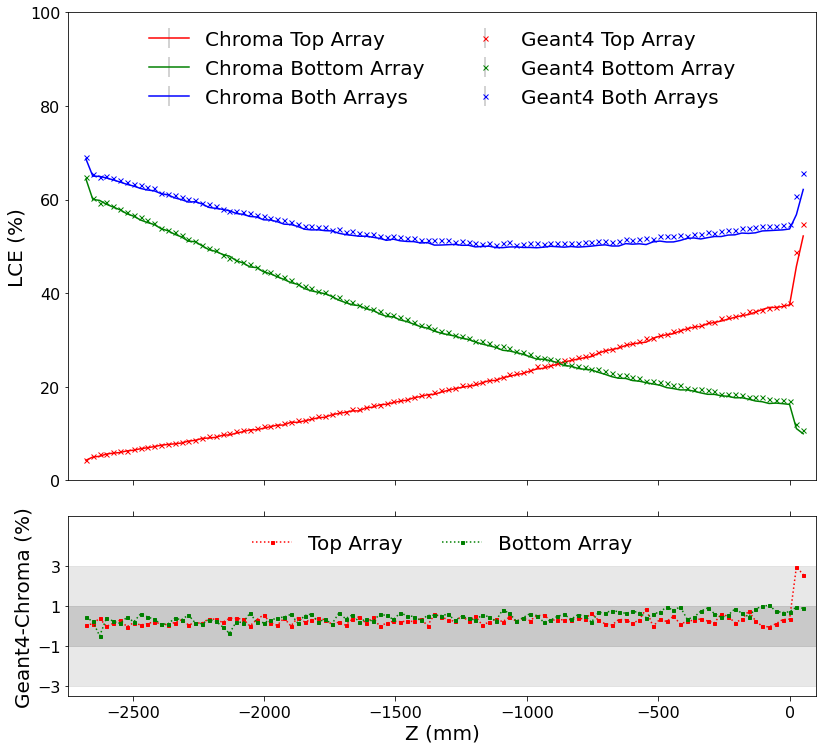}
    \caption{(Top) Total LCE as a function of the Z coordinate is shown in blue. Individual contributions of the top (bottom) PMT arrays are shown in red (green). Chroma results are shown by solid lines, while Geant4 results are shown by crosses. (Bottom) Difference between Chroma and Geant4 LCE for the top (red) and bottom (green) PMT arrays. The dark (light) gray band represents $\pm$ 1 \% abs. (3 \% abs.) deviation.}
    \label{fig:lce_vs_z}
\end{figure}
Separate contributions from the top and bottom PMT arrays are also shown. The agreement is within $\pm$1 \% abs. for all but the last top PMT array point, which is within $\pm$3 \% abs. 

To calculate the average LCE, we simulated photons with positions uniformly distributed in the active region of the detector, defined here as a volume extending from 1 mm above the cathode (Z=-2597 mm) to 1 mm below the gate (Z=-1 mm) in the Z direction and up to the PTFE sidewalls in X and Y. 10$^7$ (10$^6$) photons were emitted isotropically in Chroma (Geant4). The two frameworks were found to agree to better than 2 \% rel.. Table~\ref{tab:av_lce} shows the average LCE, total and for individual PMT arrays, for both Chroma and Geant4. The statistical uncertainty is negligible.
\begin{table}[htp]
\centering
\caption{Average LCE, total and for individual PMT arrays, determined with Chroma and Geant4. For this comparison, the electrodes were removed due to differences in implementation, the electrode frames were made unreflective, and all other parameters were set to their best guess values.}
\label{tab:av_lce}
\begin{tabular}{|l|c|c|c|}
\hline
LCE, \% & Chroma & Geant4 & Diff., \% abs. (\% rel.) \\
\hline
Top array  & 19.9  & 20.2 & 0.3 (1.5)\\
Bottom array & 32.8 &  33.2 & 0.4 (1.2) \\
Both arrays  & 52.7  & 53.4 & 0.7 (1.3)\\
\hline
\end{tabular}
\end{table}

\begin{figure}[htp]
    \centering
    \includegraphics[width=0.65\textwidth]{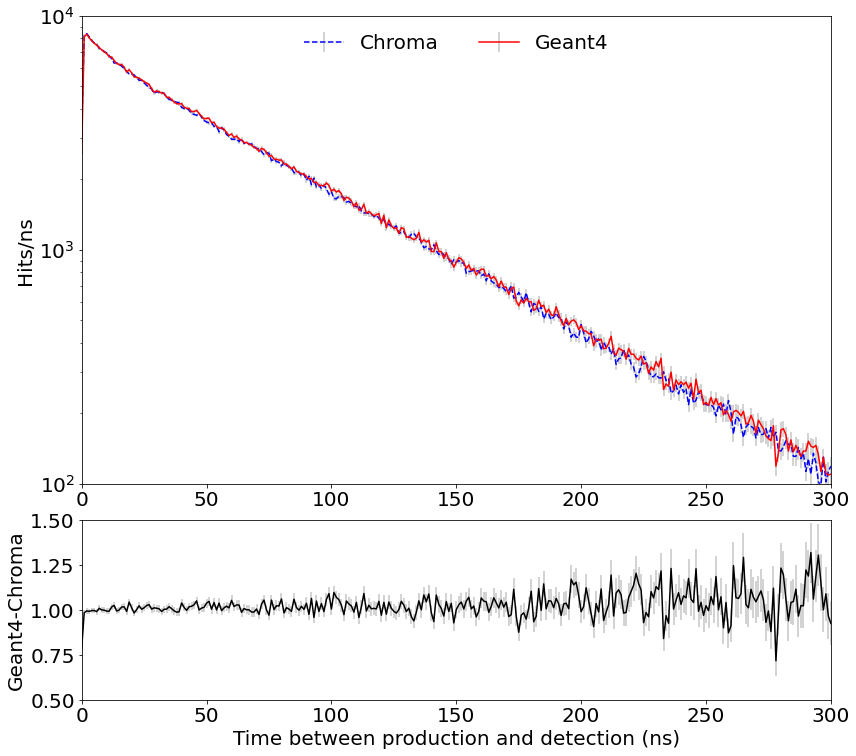}
    \caption{(Top) Distribution of time between photon generation and detection in Chroma (blue dashed line) and Geant4 (red solid line). Statistical errors are shown in grey. 10$^6$ photons were generated uniformly in the active region, using the setup of the average LCE check. (Bottom) Ratio of the Geant4 to Chroma distributions. The ratio is centered around 1, indicating a good agreement.}
    \label{fig:tcomp}
\end{figure}

The last validation check compared the photon propagation times using the simulation data generated for the average LCE check. No scintillation time profile was included in the simulation in order to only compare the effects of photon tracking, which are specific to the frameworks. Figure~\ref{fig:tcomp} shows the distributions of times between photon generation and detection for Chroma and Geant4. The ratio of the two distributions (also shown) is centered at 1, indicating a good agreement. The time to collect 90\% of detected photons, $t_{90}$, agrees to better than 2\% and is equal to 166 ns in Geant4 and 164 ns in Chroma.

The reason for the residual disagreement in the LCE and related metrics is currently unidentified but was considered small enough to be left to future work. Potential reasons are small differences in geometry or implementation of optical processes. 

As the final characterization step, the average LCE simulation was repeated using different hardware setups -- two types of GPU and three types of CPU -- and the photon tracking speed of the two frameworks was compared. Chroma was found to track photons 8 to 24 times faster than Geant4, depending on the setup. While the speedup that we observed is not as big as described elsewhere~\cite{chroma_wp, sorting}, it is possible that it becomes bigger for more complex simulations that use a larger number of sensors and geometry details. Nevertheless, the observed speedup is substantial and reduces the bottleneck associated with optical simulations. Table~\ref{tab:speedup} summarizes the photon tracking speed comparison. 
\begin{table}[htp]
\centering
\caption{Ratio of times to track the same number of photons during the average LCE simulation by Geant4 and Chroma for different hardware setups.}
\label{tab:speedup}
\begin{tabular}{|l|c|c|}
\hline
\diagbox[height=1.5\line]{\quad Geant4}{\quad \quad Chroma \quad} & GTX TITAN X~\cite{tx} & TITAN Xp~\cite{txp} \\
\hline
Intel Broadwell E5-2630v4~\cite{broadwell}  & 11 & 16 \\
Intel Xeon Silver 4214~\cite{xeon} & 8 & 12 \\
AMD Opteron 6220~\cite{amd} & 17 & 24 \\
\hline
\end{tabular}
\end{table}

We considered the results of Chroma's validation and characterization to be adequate and proceeded using the framework.

\section{Baseline simulation study with Chroma}

\subsection{Light collection efficiency}
We calculated the average LCE for the baseline detector geometry using Chroma. The uncertainty of the optical parameters summarized in Table~\ref{tab:optical} leads to a systematic spread of the average LCE values. We estimated the spread due to the main optical parameters by performing the average LCE simulation with each of the optical parameters adjusted one by one to its assumed minimum and maximum value. The results are summarized in Table~\ref{tab:lce_sys}, reiterating the known importance of choosing highly reflective PTFE and achieving good LXe  absorption length. The possibility of achieving the LXe absorption length even larger than the one assumed as the upper limit in Table~\ref{tab:optical} is considered in the next subsection. 
\begin{table}[htpb]
\centering
\caption{Average LCE corresponding to different values of the optical parameters. One parameter is changed at a time for this study, with the range of the variation corresponding to its assumed systematic spread.}
\label{tab:lce_sys}
\begin{tabular}{|l|c|c|}
\hline
Parameter & Spread & LCE, \%\\
\hline
LXe L$_{\mathrm{abs}}$, m  & 20 -- 100 & 32 -- 49 \\
LXe L$_{\mathrm{Rayleigh}}$, cm  & 20 -- 50 & 40 -- 45\\
LXe $R_{\mathrm{total}}^{\mathrm{PTFE}}$, \%  & 93 -- 100 & 34 -- 45\\
\hline
\end{tabular}
\end{table}
The average LCE obtained with all optical parameters at their best guess values (Table~\ref{tab:optical}) is 43\%, with 15\% (28\%) 
detected by the top (bottom) PMT array. An average LCE of 36\% has been reported for the XENONnT detector~\cite{XENONnT}. While not all relevant parameters could be found in that work, a smaller LCE is expected given the larger-diameter, unreflective wire electrodes used in the XENONnT simulation. 

\begin{figure}[htpb]
    \centering
    \includegraphics[width=0.49\textwidth]{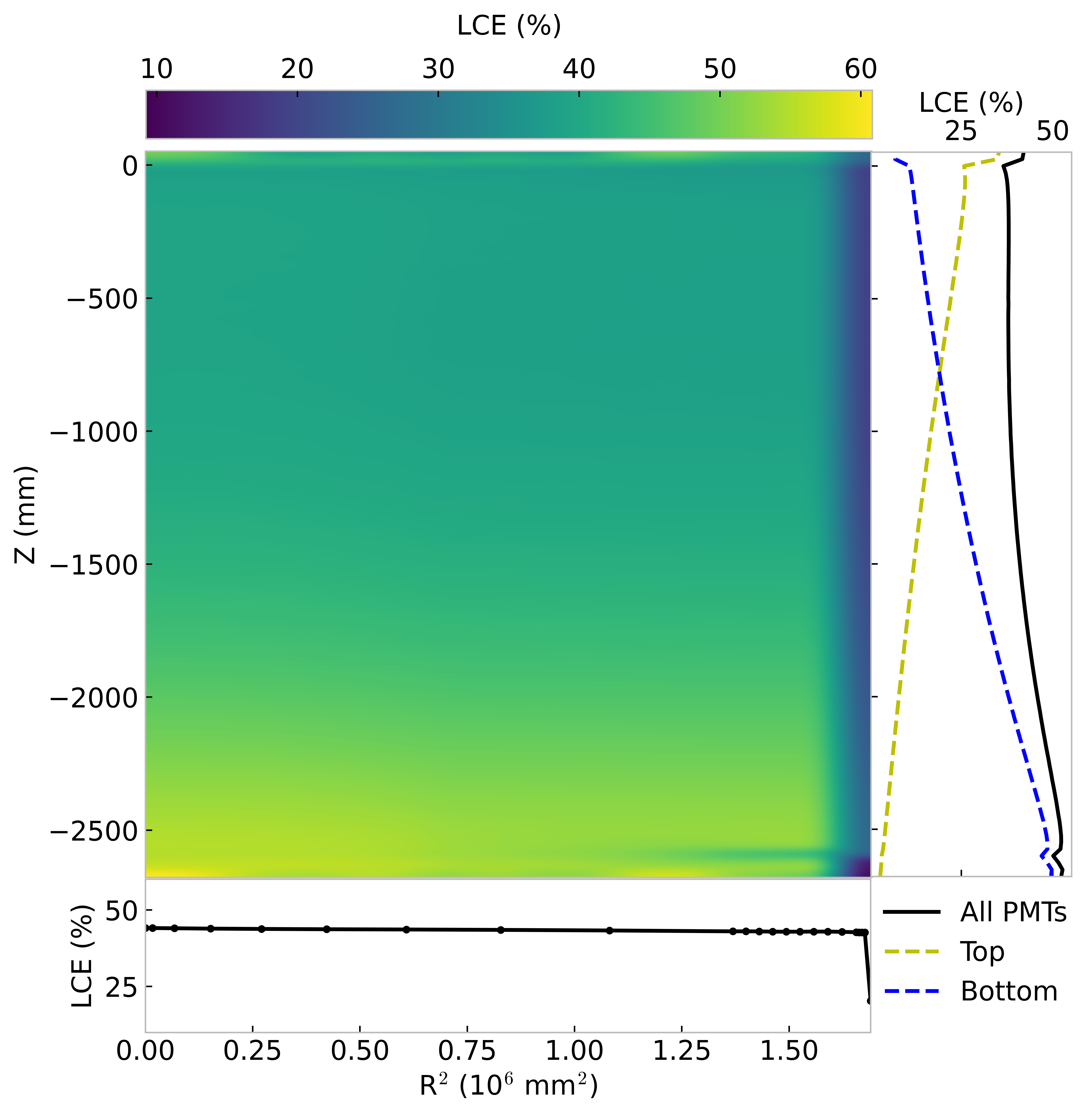}
    \caption{Dependence of LCE on Z and R$^2$. Projections on the Z and R$^2$ axes are shown by solid lines on the right and bottom panels. Yellow and blue dashed lines show the Z dependence of the LCE for the top and bottom PMT arrays, respectively. The point corresponding to the same fraction of photons detected by the two arrays is closer to the top PMT array. The dip near the bottom edge of the detector is due to proximity to the electrode frame, which has a substantially lower reflectivity than PTFE. The effect is not present near the top, due to the top electrode frames being covered with a PTFE reflector.}
    \label{fig:2d}
\end{figure}

To study the spatial variation of the LCE, the simulations were performed with isotropic, point light sources distributed on a 2-D grid. Since negligible $\phi$ dependence is expected due to symmetry, the radial direction was probed by increasing the Y coordinate while keeping the X coordinate at zero. Close to 2500 simulations were performed on a ZxR$^2$ grid with 10$^6$ photons generated at each position. Figure~\ref{fig:2d} shows the results. The LCE shows a noticeable variation along the Z direction, with the point corresponding to the same fraction of photons detected by the two arrays being closer to the top PMT array. The LCE remains stable to within 1.5\% radially, up until the last centimeter to the edge of the detector. 
The dip in the LCE near the bottom edge of the detector is due to the proximity to the bottom electrodes' SS frames, which have a substantially lower reflectivity than PTFE. The effect is not present near the gate, due to the top frames being covered with a PTFE reflector.

\begin{figure}[htpb]
    \centering
    \includegraphics[width=0.49\textwidth]{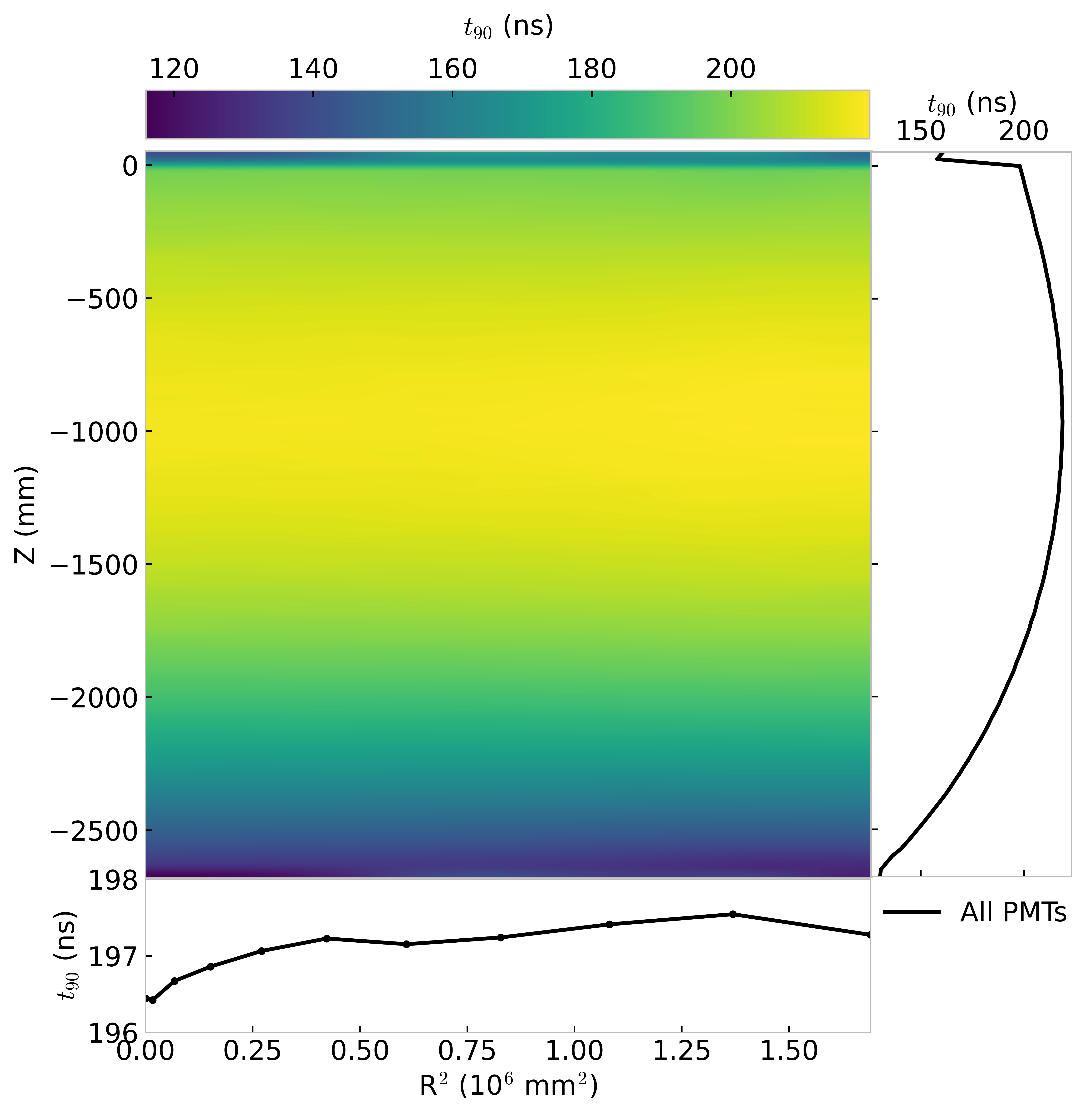}
    \caption{Dependence of $t_{90}$ on Z and R$^2$. Photons originating closer to the PMT arrays take less time to get detected, on average. Only a minor radial dependence is seen, as expected. The average $t_{90}$ is around 200 ns, which is much larger than the time it takes a photon to traverse the full height of the detector and is a consequence of short scattering length.}
    \label{fig:2dt}
\end{figure}

Another important factor in the design of the next-generation detector is the time it takes to collect the detected optical photons. The narrower the distribution of photon arrival times is, the smaller the rate of dark noise coincidences. We used the results of the above simulations to calculate the time it takes to collect 90\% of detected photons. The simulations included the scintillation time profile for NR events, but using the $f_{\mathrm{s}}$ constant for ER events leads to only a 2 ns change in $t_{90}$. The average $t_{90}$ is around 200 ns. The spatial variation of $t_{90}$ is shown in Figure~\ref{fig:2dt} as a function of Z and R$^2$. 
The average $t_{90}$ is more than an order of magnitude larger than the time it takes a photon to traverse the full height of the DARWIN detector without any interaction. This is primarily a consequence of the fact that the Rayleigh scattering length is much smaller than the size of the detector (see Table~\ref{tab:optical}). The wide distribution of the scattering angles (Figure~\ref{fig:ray_comparison}) leads to photons taking a much longer time to reach the photodetectors. A similar effect is the diffuse reflection off the PTFE sidewall. However, the distance between the PTFE panels in DARWIN is much larger than the scattering length. 

\subsection{Sources of photon losses}
\label{sec:graveyards}
\begin{figure}[htpb]
    \centering
    \includegraphics[width=0.49\textwidth]{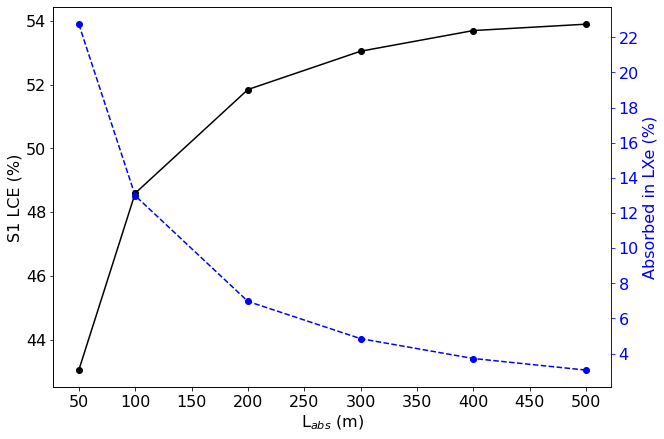}
    \caption{Dependence of the average LCE (black points, solid line) and fraction of photons absorbed by LXe (blue points, dashed line) on the LXe absorption length used in the simulation. The LCE approaches the asymptotic value of $\sim$54\% for L$_{\mathrm{abs}}$ of 400-500 m.}
    \label{fig:labs}
\end{figure}

One way to understand what design choices could lead to the largest impact on LCE is to look at where and how the photons are absorbed in the detector. In our simulation, out of the 56.9\% of photons that are produced in the active region and not detected, the largest fraction, $\sim$22\%, gets absorbed in LXe. The LXe absorption length assumed in the simulation is 50 m, which is based on a conservative assumption of a current-generation experiment~\cite{XENONnT}. Absorption in LXe is caused by impurities, most notably water and oxygen~\cite{baldini}. It is conceivable that the next-generation detector will be able to achieve a substantially better LXe purity~\cite{aprile_10ms}, leading to a larger LXe absorption length. Figure~\ref{fig:labs} shows the potential impact of improved absorption length on the average LCE up to the optimistic assumption of a ten-fold improvement in L$_{\mathrm{abs}}$.
The figure shows that for absorption lengths above 400-500 m, the average LCE approaches the asymptotic value of $\sim$54\%, which is a substantial improvement over the best guess value assumed in the baseline design.

The second-largest fraction, $\sim$18\%, is absorbed by the electrodes. The gate electrode absorbs more than 7\% of photons, apparently due to it being positioned just below the LXe/GXe interface that leads to some photons passing through it several times. This suggests that it would be advantageous to make the gate electrode as transparent as possible. Other electrodes and frames absorb from a fraction of a percent up to 2-3\% each. 

The PTFE sidewall reflector absorbs about 7\% of photons. The rest of photon absorption happens inside the PMTs: roughly 6\% (3\%) of photons are entering the quartz window but miss the photocathode in the bottom (top) PMT array. The higher number of photons absorbed by the bottom array reflects its larger contribution to the total LCE, as compared to the top one, such that the ratio of absorbed to detected photons is roughly the same for both arrays. Table~\ref{tab:graveyard} summarizes the results.
\begin{table}[htpb]
\centering
\caption{Sources of photon losses in a DARWIN-like detector.}
\label{tab:graveyard}
\begin{tabular}{|l|c|}
\hline
Component & Absorbed photons, \% \\
\hline
\textbf{LXe ($L_{\mathrm{abs}}$ = 50 m)} &\hspace{0pt} \textbf{22.7} \\
\textbf{Electrodes} &\hspace{0pt} \textbf{18.0} \\
\quad Gate &\hspace{1ex} 7.7 \\
\quad Cathode &\hspace{1ex} 3.1 \\
\quad Bottom screen &\hspace{1ex} 2.6 \\
\quad Anode &\hspace{1ex} 1.7  \\
\quad Top screen &\hspace{1ex} 1.6  \\
\quad Frames &\hspace{1ex} 1.3 \\
\textbf{PMTs} &\hspace{1ex} \textbf{9.0} \\
\quad Bottom PMTs &\hspace{1ex} 6.3 \\
\quad Top PMTs    &\hspace{1ex} 2.7 \\
\textbf{PTFE reflectors} &\hspace{1ex} \textbf{7.2} \\
\hline
\textbf{Total} &\hspace{0pt} \textbf{56.9}\\
\hline
\end{tabular}
\end{table}

\section{Alternative designs}
\subsection{Reflective coatings}
Table~\ref{tab:graveyard} shows that electrodes absorb a large fraction of photons, with the largest absorption at the gate electrode. Reducing photon losses by the electrodes is possible by optimizing the design of the electrodes to achieve higher transparency. For example, one could consider using thinner wires with larger spacing, or by substituting some or all wire electrodes with thin transparent disks covered by conductive, transparent films. Alternatively, one could retain the current mechanical design but increase the reflectivity of the electrodes' wires. Here, we pursue the latter option. A relatively common way to create a highly reflective VUV mirror is to coat a material with thin films of Al and MgF$_2$. Aluminum is known to be highly reflective in the VUV region. MgF$_2$ is transparent in the VUV and serves as a protective layer that prevents oxidation of the aluminum. 

Based on standalone theoretical calculations of Fresnel reflections in thin films and experience with commercially available Al/MgF$_2$ mirrors~\cite{pelham}, we assume that a coating can achieve 90\% total reflectivity. The theoretical calculations predict $\sim$10\%-15\% variation of reflectivity with the angle of incidence after taking into account the wavelength distribution of the LXe scintillation. We ignore this effect for simplicity in this study. Notably, as mentioned in Section~\ref{sec:optical}, theoretical calculations do not always accurately predict reflectivity at VUV wavelengths and in LXe, so a dedicated measurement would be warranted if the simulation results are promising. Similarly, we consider practical aspects of incorporating reflective coatings in the electrode design, namely potential charge accumulation, spurious electron emission, background contribution, and mechanical stability, to be outside the scope of this study. These aspects do not appear to be showstoppers but would require a careful consideration if this design option is adopted by the experiment.

To quantify the potential improvement to the LCE from covering the electrodes and their frames with Al/MgF$_2$, we assigned 90\% reflectivity to the corresponding components and repeated the simulation with 10$^6$ photons generated uniformly in the active region of the detector. All other parameters and optical properties were kept at their best guess values. A separate case of only covering the gate electrode was considered first. Since the gate absorbs several times more photons than other electrodes, covering only the gate could be a compromise solution in case radiopurity or other tests suggest that covering all electrodes is too risky or too expensive. Table~\ref{tab:results} shows the average LCE, including separate contributions from the top and bottom PMT arrays, for the considered scenarios. 
\begin{table*}[htpb]
\centering
\caption{Average LCE with and without reflective coatings on the electrodes. The coatings are assumed to be 90\% reflective. Improvements for the best guest and optimistic values of the LXe absorption length are shown.}
\label{tab:results}
\begin{adjustbox}{width=1\textwidth}
\begin{tabular}{|c|l|c|c|}
\hline
LXe L$_{\mathrm{abs}}$, m & Scenario & Total LCE (top/bottom), \% & Improvement, \% abs. (\% rel.)\\
\hline
\multirow{3}*{50} & Baseline (SS electrodes+frames) & \textbf{43.1} (14.8/28.2) & -\\ 
                  & Only Gate covered with Al/MgF$_2$ & \textbf{46.0} (16.8/29.2) & +2.9 (+6.7) \\
                  & All electrodes and frames covered & \textbf{51.8} (18.6/33.2) &  +8.7 (+20)\\ 
\hline
\multirow{3}*{500} & Baseline (SS electrodes+frames) & \textbf{54.0} (18.7/35.3) & - \\ 
                  & Only Gate covered with Al/MgF$_2$ & \textbf{58.7} (21.8/36.9) & +4.7 (+8.5) \\
                  & All electrodes and frames covered & \textbf{67.0} (24.3/42.7) &  +13 (+24)\\ 
\hline
\end{tabular}
\end{adjustbox}
\end{table*}

For a perfect film on a perfectly smooth substrate, all reflections are expected to be specular. Roughness of the substrate/film leads to an increase in diffuse reflections. Based on experience with commercial Al/MgF$_2$ mirrors, we assume that most reflections are specular (85\% out of the 90\% total). The results shown in Table~\ref{tab:results} are quite stable with respect to the specular/diffuse ratio. In particular, making the opposite assumption -- 85\% out of 90\% total reflection is diffuse -- leads to only a $\sim$0.2\% change in the LCE.

While the default simulation assumes the electrodes to be made from parallel wires of 200 \si{\micro}m diameter, we checked that the conclusions of this section are stable against alternative electrode designs. To that end, CAD models of electrodes consisting of hexagonal meshes with 10.2 mm and 3.5 mm cell opening and 178 \si{\micro}m wire diameter were made. Two scenarios were then investigated. First, all five detector electrodes were assumed to be meshes with 178 \si{\micro}m diameter and 10.2 mm cell opening. Second, only the gate electrode was assumed to be a mesh (with a 3.5 mm cell opening). In both cases, a similar improvement in the LCE was seen when the electrodes were covered with the reflective coating (51\% coated vs. 43\% uncoated for the first scenario and 45\% vs. 40\% for the second one).

Coating the electrodes with reflective films leads to even larger relative improvement in LCE if one simultaneously improves the LXe absorption length. Table~\ref{tab:results} shows the LCE improvement from coating the electrodes for the optimistic case of 500~m absorption length. It should be noted, however, that this would also increase $t_{90}$ from $\sim$200 to $\sim$240 ns. We conclude that covering the electrode with a reflective coating is worth serious consideration for DARWIN.

\subsection{4$\pi$ design}
\label{sec:4pi}
As a potential improvement over the baseline design, we investigated placing extra photosensors around the sides of the detector. The goal is to achieve as complete coverage as practical, hence the ``4$\pi$'' designation. Two choices of photosensors for the side array were considered, Hamamatsu's S13371 VUV4 SiPMs~\cite{vuv4} and Hamamatsu's flat, square 2-inch R12699-406-M4 PMTs~\cite{2in}.

\begin{figure}[htpb]
    \centering
    \includegraphics[width=0.95\textwidth]{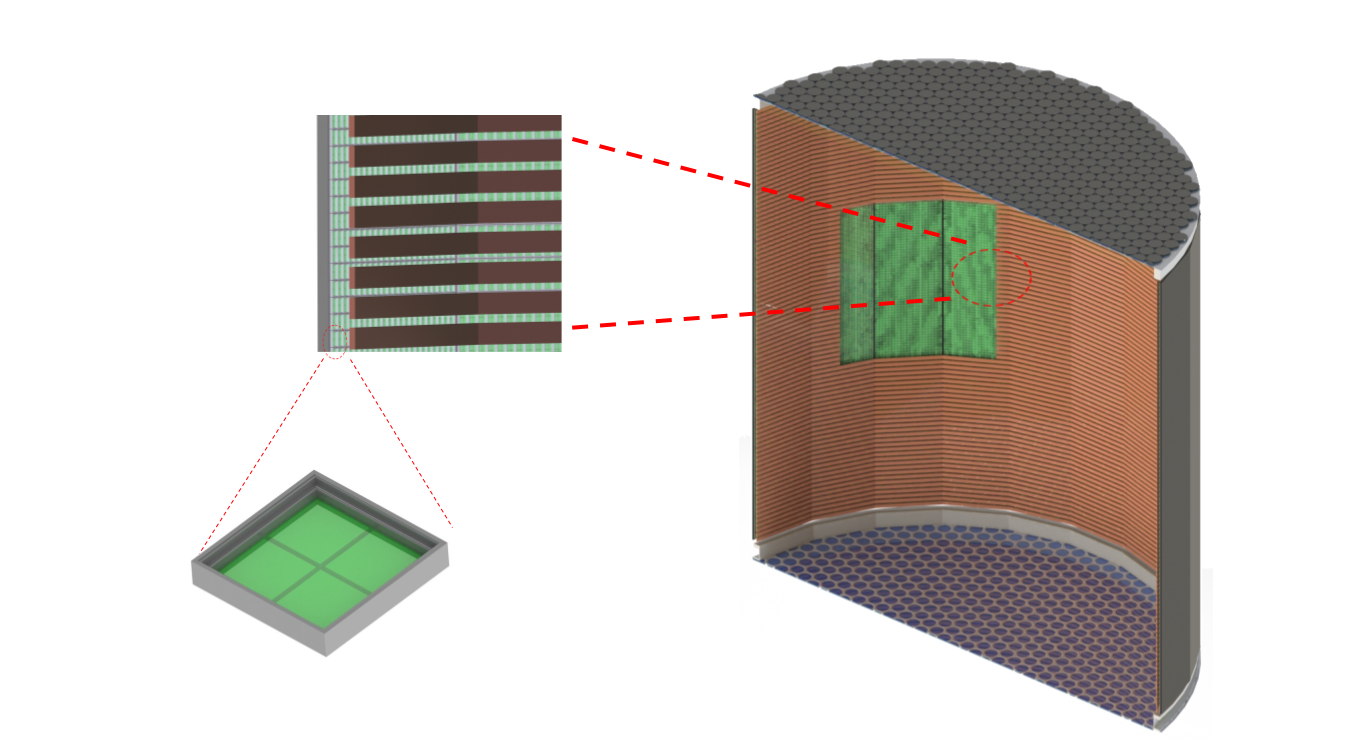}
    \caption{Cut-out view of the 4$\pi$ design variant. The FSRs are shown in gold. A cutout of the FSRs reveals the SiPMs (green). The photocathodes of the bottom PMT array are seen as blue disks. The backing plates and photon-absorbing caps of the top PMT array are seen in dark grey. The PTFE reflectors around the PMTs and electrode frames are shown in light grey. Close-up views shows the backing plates (dark grey) with SiPMs positioned behind the FSRs and an individual VUV4 Hamamatsu quad SiPM. The SiPM is shown with the ceramic holder (light gray) that we assume to not reflect any light. The translucent green represents the reflective film in front of the SiPM's sensitive area. The film completely covers the active area, shown in dark green. }
    \label{fig:4pi_Section}
\end{figure}

SiPMs are potentially attractive photosensors for the next-generation LXe detectors~\cite{sipm_1,sipm_2}, provided the dark count rate is further reduced so as to not affect the threshold properties needed for the WIMP search. One possible use for SiPMs is replacing the PMTs altogether, which has the advantage of reducing the radioactive backgrounds. The alternative approach considered here is to keep the existing top and bottom PMT arrays in place and add SiPMs along the barrel of the TPC. To simulate this scenario, we created a model of a SiPM array, removed the side PTFE reflector panels, and added the FSRs. The baseline DARWIN design contains 92 copper FSRs placed immediately behind the side PTFE reflector panels with a vertical step size of 28 mm. Each FSR has 20 mm height and 5 mm width. The vertical step size is defined as the distance between the centers of two consecutive FSRs, so there are only 8 mm between the edges of each two nearby rings. Individual SiPMs were modeled after the VUV4 Hamamatsu quad SiPM, which has an active area of 5.85 $\times$ 5.95 $\times$ 4 mm$^2$ and comes integrated in a 15 $\times$ 15 mm$^2$ holder. A thin 1.5 \si{\micro}m layer was added immediately in front of the active layer to allow modeling of the SiPM reflectivity separately from the SiPM detection efficiency. Individual quad SiPMs are tiled onto backing plates, which provide mechanical support and electrical connections to the sensors. Each of the backing plates has dimensions of 350 $\times$ 2568 mm$^2$ and contains 22 $\times$ 165 of individual SiPMs tiled onto it with a spacing of 0.5 mm separating them from one another. There are 24 plates with a total of more than 87000 individual quad SiPMs in the 4$\pi$ design. The total photosensitive area of all SiPMs is 12.1 m$^2$, but roughly half of this area is obstructed by the FSRs. 
The plates are placed behind the FSRs such that the top of the panels is parallel to the top of the FSRs and the bottom is level with the bottom of the FSRs. This constrains the SiPMs to the LXe region of the detector. The front of the holders is placed 17 mm from the FSRs. This preliminary placement was chosen so that the electrical field created by the FSRs is not expected to interfere with the SiPMs~\cite{efield_sipm}. This placement gives a distance from the active area to the FSRs of 18.6 mm. A more detailed look into this matter must be completed before a final placement can be chosen. Figure~\ref{fig:4pi_Section} shows an overall cross-sectional view of the design variant. In the figure, one can also see the individual SiPMs that are tiled onto the backing plate.

The Hamamatsu 2-inch R12699-406-M4 flat PMT~\cite{2in} is another potentially attractive photosensor for use in the side array. To simulate this scenario, the same approach as in the SiPM variant was followed. Individual PMTs have a square active area of 48.5 $\times$ 48.5 mm$^2$ and come integrated in a 56 $\times$ 56 mm$^2$ body. A 2.5 mm quartz window is placed immediately in front of the active surface. The PMT body is assumed to not reflect any light. Each of the backing plates for the 2-inch PMT variant contains 6 $\times$ 45 of individual PMTs tiled onto it with a spacing of 0.5 mm separating them from one another. The total number of PMTs is close to 6500, giving an active area of 15.2 m$^2$. The front of the PMT's envelope is placed 18.6 mm from the FSRs, while the quartz window is 16.1 mm from the FSRs. This preliminary placement was chosen so that the active surface of the PMTs is placed at the same distance away from the FSRs as in the SiPM variant.

Table~\ref{tab:optical_4pi} shows the optical properties assumed for the materials relevant for the 4$\pi$ design variant simulation and not already specified in Table~\ref{tab:optical}. 
\begin{table*}[htpb]
\centering
\caption{Optical properties of materials relevant to the 4$\pi$ design.}
\label{tab:optical_4pi}
\begin{adjustbox}{width=1\textwidth}
\begin{tabular}{|l|c|c|c|}
\hline
Parameter & Best Guess Value @175 nm & Spread & Components\\
\hline
Copper reflectivity, $R_{\mathrm{total}}$ \% & 45 &  & \multirow{2}*{FSRs} \\
\cline{1-3}
Specular/diffuse reflectivity ratio for copper, $R_{\mathrm{spec}}$/$R_{\mathrm{diff}}$ & 3.5 &  &  \\
\hline
Aluminum reflectivity, $R_{\mathrm{total}}$ \% & 90 & 90 -- 98 & \multirow{2}*{Reflective coating} \\
\cline{1-3}
Specular/diffuse reflectivity ratio for aluminum, $R_{\mathrm{spec}}$/$R_{\mathrm{diff}}$ & 17 & 17 -- 99 &  \\
\hline
SiPM reflectivity, $R_{\mathrm{spec}}$ \% & 25~\cite{lixo} & 20 -- 28~\cite{lixo} & SiPM film\\
\hline
Silicon refractive index, $n_{\mathrm{Silicon}}$  & 0.81~\cite{SiData_2} &  & \multirow{2}*{Backging plates} \\
\cline{1-3}
Silicon extinction coefficient, $k_{\mathrm{Silicon}}$ & 1.86~\cite{SiData_2} &  & \\
\hline
Ceramic reflectivity, $R_{\mathrm{total}}$ \% & 0 & & SiPM holders\\
\hline
\end{tabular}
\end{adjustbox}
\end{table*}

To quantify the LCE of the 4$\pi$ design, 10$^6$ photons were simulated uniformly in the active region of the detector. For all simulations, the total average LCE, as well as individual contributions from the top, bottom, and side arrays, were determined. For the material of the backing plates, we investigated two options -- quartz and silicon -- that gave results consistent within the statistical error, which is expected given the close packing of the photosensors on the plates. Ultimately, we set the backing plate optical properties to full absorption, as the most conservative case. We considered two other factors in this investigation. Firstly, we considered coating the copper FSRs with a reflective film, such as the Al/MgF$_2$ coating investigated earlier. While adding to the cost and technical challenges, coating the FSRs has a positive impact on the LCE, which in case of the 2-inch PMT variant slightly surpasses the baseline design. Secondly, we show the case of the DARWIN detector without FSRs at all, to illustrate the absolute maximum improvement from the 4$\pi$ design. 
Table~\ref{tab:4pi_sipm} summarizes the results. 
\begin{table}[htpb]
\centering
\caption{LCE for the 4$\pi$ variants with fully-absorbing backing plates. All optical parameters set to their best guess values.}
\label{tab:4pi_sipm}
\begin{adjustbox}{width=0.49\textwidth}
\begin{tabular}{|l|c|c|}
\hline
Variant & FSRs & Total LCE (top/bottom/side), \% \\
\hline
Baseline & - & \textbf{43.1} (14.8/28.2/-)\\
\hline
\multirow{3}*{SiPMs} & Copper & \textbf{22.6} (4.3/9.3/9.1) \\
                          & Al/MgF$_2$ & \hspace{2ex}\textbf{36.9} (6.1/13.0/17.9) \\
                          & No FSRs & \hspace{1ex}\textbf{44.4} (3.7/8.1/32.6)\\
\hline
\multirow{3}*{PMTs} & Copper & \hspace{1ex}\textbf{27.7} (4.3/9.3/14.1) \\
                           & Al/MgF$_2$ & \hspace{2ex}\textbf{46.8} (6.0/13.0/27.8) \\
                           & No FSRs & \hspace{1ex}\textbf{61.2} (3.5/7.8/49.9) \\
\hline
\end{tabular}
\end{adjustbox}
\end{table}

It is clear from the table that the FSRs are a major issue for the 4$\pi$ design, decreasing the potential advantage of placing the photosensors behind them. In both variants, the top and bottom arrays detect a substantially smaller fraction of photons, as compared to the baseline design (13.6\% for the both 4$\pi$ variants with copper FSRs, compared to 43.1\% for the baseline design). In the case of the 2-inch PMT variant the side array becomes the dominant detector of light, surpassing the bottom PMT array. The 2-inch PMT variant leads to a higher LCE than the VUV4 SiPM one due to the larger total active area and less reflection from the entrance layer. If one can reduce the thickness of individual FSRs and/or increase the spacing between them, the advantage of the 4$\pi$ approach will increase accordingly, but only for the 2-inch PMT variant the LCE is expected to surpass the baseline design. Changing the FSRs needs to be carefully weighted against potential drawback of reduced uniformity of the electric field. Another issue is the dark noise rate associated with the large added active area, which can be a problem even when using PMTs as photosensors. 

\begin{figure}[htpb]
    \centering
    \includegraphics[width=0.49\textwidth]{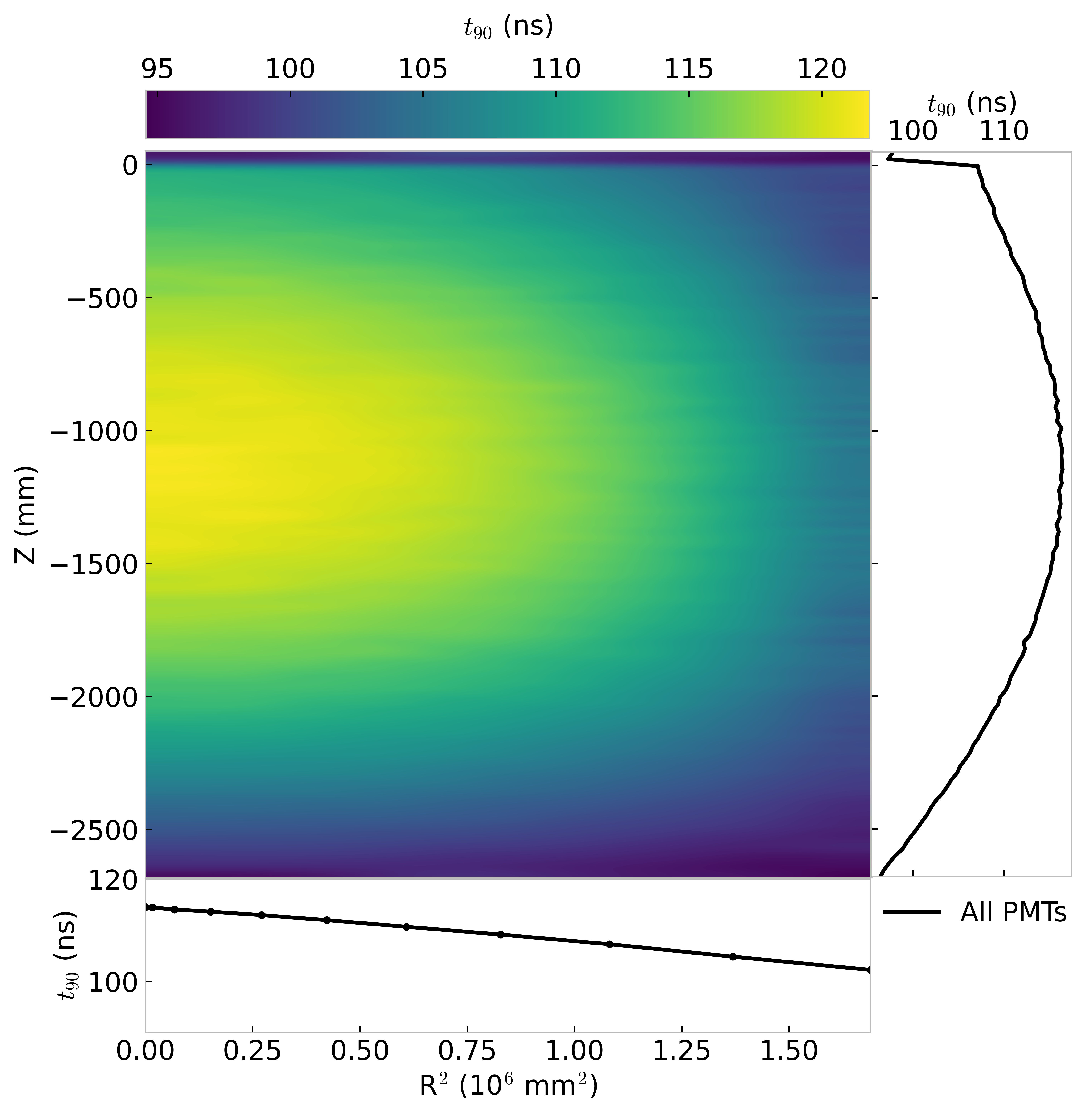}
    \caption{$t_{90}$ as a function of Z and R$^2$ position in the detector for the 4$\pi$ design variant. The average $t_{90}$ is 110 ns, consistent to within 1 ns between the VUV4 SiPM and 2-inch PMT design variants. The decrease with larger radius is due to the proximity to the side photosensor array.}
    \label{fig:4pi_2dt}
\end{figure}

Figure~\ref{fig:4pi_2dt} shows the $t_{90}$ as a function of position in the detector. The average value of $t_{90}$ is around 110 ns for both VUV4 SiPM and 2-inch PMT variants, compared to about 200 ns for the baseline design. Qualitatively, the reduction is expected given a more direct path to the photosensors in the absence of the sidewall PTFE reflector. The general features of the 2D distribution of $t_{90}$ are also intuitive. In particular, the noticeable decrease with larger radius is due to the proximity to the side photosensor array. 

\begin{figure}[htpb]
    \centering
    \includegraphics[width=0.49\textwidth]{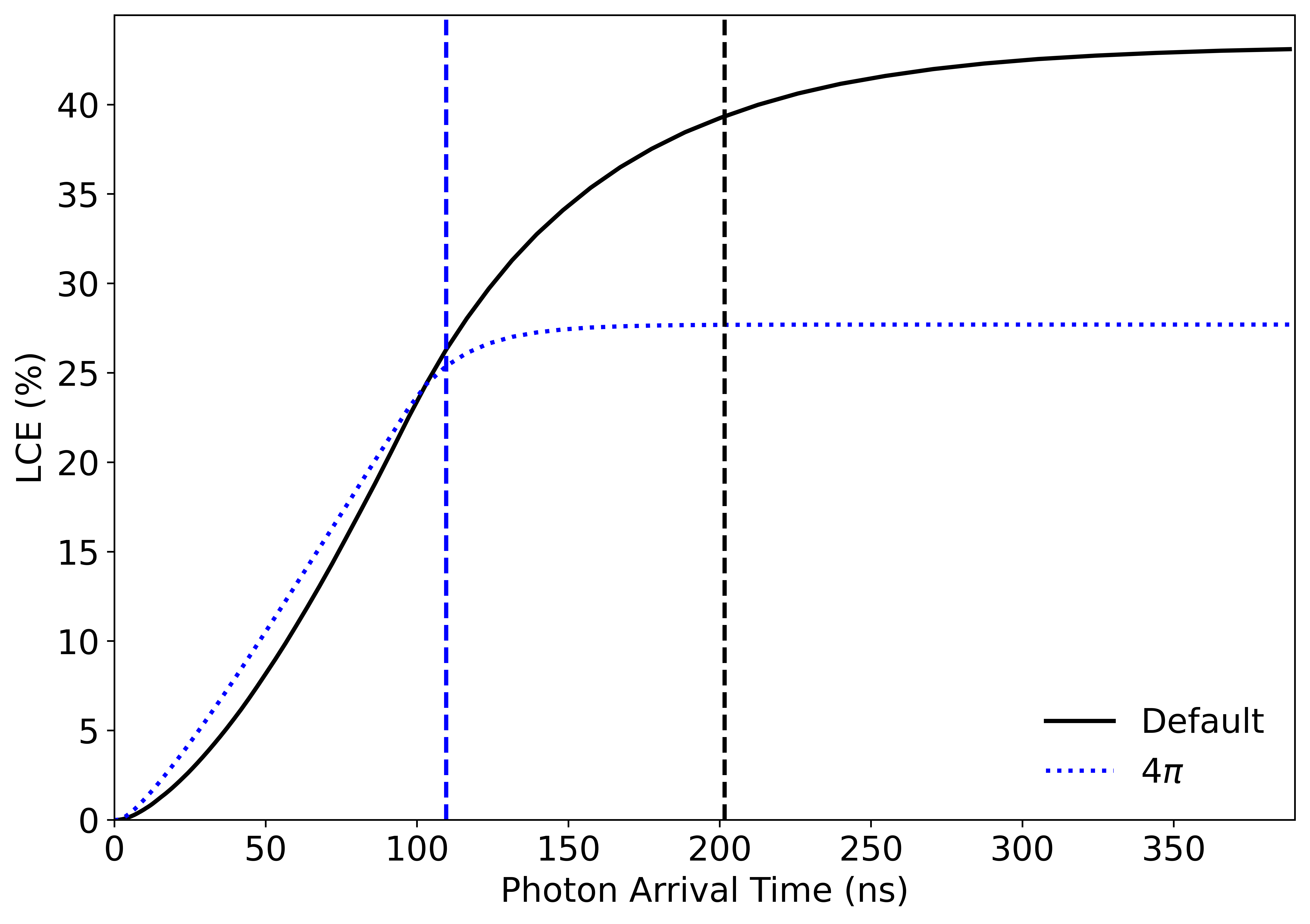}
    \caption{The cumulative distribution of photon arrival times averaged over the active region in the 2-inch PMT variant of the 4$\pi$ design (blue dotted line) and the baseline design (black solid line). The dashed vertical lines mark the time it takes to collect 90\% of detected photons, $t_{90}$. The two curves are normalized by the average LCE values.}
    \label{fig:4pi_td}
\end{figure}

Figure~\ref{fig:4pi_td} shows the cumulative distribution of photon arrival times averaged over the active region in the 4$\pi$ and baseline designs. The two curves are normalized by their respective average LCE values, which emphasizes that the faster arrival time in the 4$\pi$ variant is offset by the smaller LCE. 
Interestingly, as can be deduced from the figure, if one wanted to place a cut on the collection time in the baseline design at the $t_{90}$ value of the 4$\pi$ variant, then the effective average LCE of the baseline design would become close to the average LCE of the 4$\pi$ variant. While such a cut would be wasteful, it shows that LCE or $t_{90}$ by themselves are not sufficient metrics for optimizing the design of the experiment. All aspects affecting the physics reach need to be considered simultaneously. 

\subsection{Single-phase TPC}
\begin{figure}[htpb]
    \centering
    \includegraphics[width=0.65\textwidth]{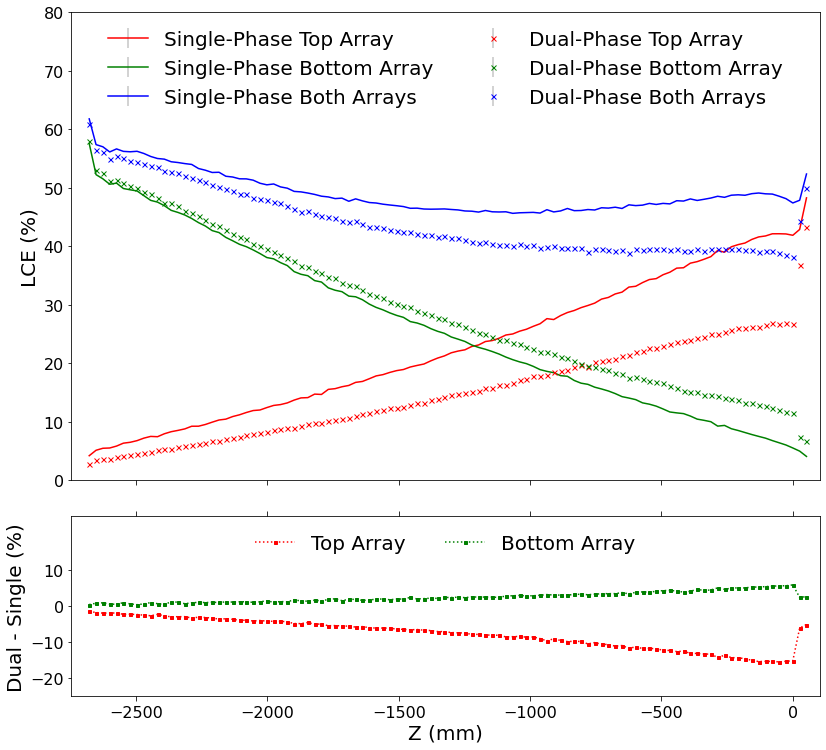}
    \caption{(Top) Total LCE as a function of the Z coordinate is shown in blue. Individual contributions of the top (bottom) PMT arrays are shown in red (green). Results for the single-phase design are shown by solid lines, while the baseline design results are shown by crosses. (Bottom) Difference between the baseline and single-phase LCE for the top (red) and bottom (green) PMT arrays.}
    \label{fig:phases}
\end{figure}

Electroluminescence in LXe~\cite{sp_1,sp_2} received renewed interest in recent years~\cite{aprile_1,sp_3,sp_4,radial_tpc,single_phase} and offers several potential advantages for the next-generation LXe detector~\cite{sp_4,single_phase}. Among possible ways this approach may be realized, the simplest one is by removing the LXe/GXe boundary in the baseline detector design, such that LXe fills up to the top PMT array, and increasing the field strength between the gate and anode electrodes above the electroluminescence threshold. This, minimal, modification is expected to slightly underestimate the potential LCE improvement, as the requirement of thinner and wider-spaced anode wires in a single-phase detector will further reduce photon losses. To evaluate the impact of this design choice on light collection, we adjusted the baseline design accordingly and evaluated LCE at different Z positions in the detector. 10$^5$ photons were generated isotropically at $\vec{r}$=(0,0,Z), where Z varied from -2648 to 52 mm. Figure~\ref{fig:phases} shows the Z dependence of the LCE for the single-phase and baseline designs. Separate contributions from the top and bottom PMT arrays are also shown. 
The figure shows that the strong bottom-top asymmetry present in the baseline design is decreased in the single-phase variant. The average LCE of this design increases by 4.9 \% abs. (11 \% rel.). Aside from the overall LCE increase, the lack of reflection on the LXe/GXe interface reduces the mean photon propagation time. The time to collect 90\% of the detected photons decreases from about \SI{200}{ns} in the baseline design to \SI{175}{ns} in the single-phase design.

\subsection{No bottom PMT array}
\begin{figure}[htpb]
    \centering
    \includegraphics[width=0.49\textwidth]{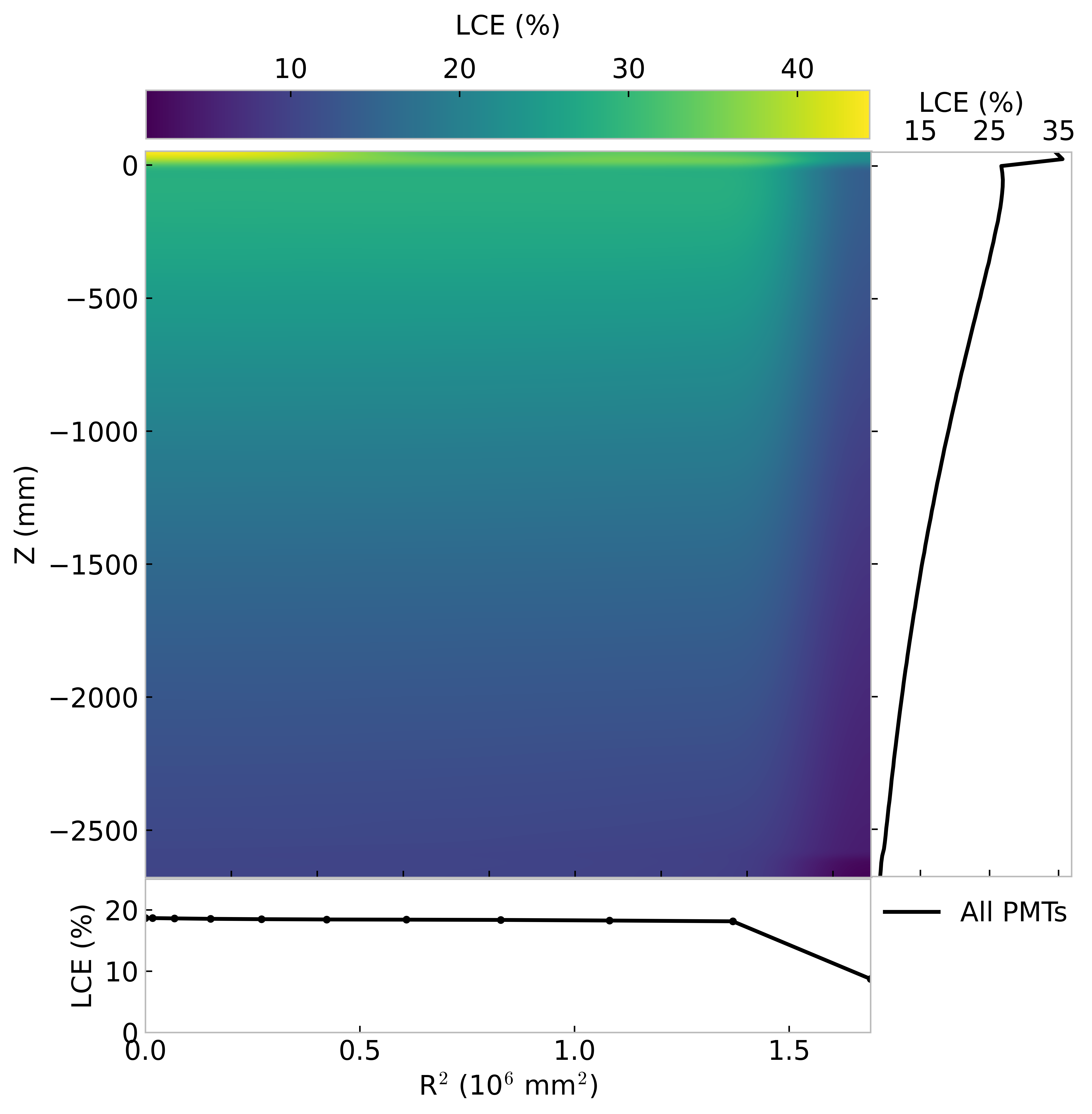}
    \caption{Dependence of LCE on Z and R$^2$ for the design variant without the bottom PMT array. Projections on the Z and R$^2$ axes are shown by solid lines on the right and bottom panels. A factor of $\sim$3 smaller LCE occurs close to the bottom of the detector due to the long optical path to photosensors.}
    \label{fig:no_bot}
\end{figure}

For the last example study we adjusted the detector design by removing the bottom PMT array and replacing it with a PTFE reflective film. One might argue that this approach may retain a reasonable LCE while decreasing the number of photosensors and related electronics by half, hence reducing the backgrounds and costs. However, as can be seen from Figure~\ref{fig:no_bot}, the reduction in LCE in a large detector like DARWIN is too dramatic in spite of the high reflectivity of the bottom PTFE reflector (99\%). 

The LCE's Z dependence is now stronger, with a factor of $\sim$3 smaller LCE occurring close to the bottom of the detector, which is expected given the long optical path to photosensors. The LCE averaged over the active region is 18\%, less than a half of the baseline design value. Increasing the LXe absorption length from 50 to 500 m increases the average LCE by $\sim$7.5\%, such that it still lags far behind the baseline design variant. The time to collect 90\% of detected photons also becomes less attractive in this design variant, increasing to 265 ns. We conclude that it would be difficult to justify removing the bottom PMT array in the DARWIN detector. 

\section{Summary and Conclusions}
In this work we investigated an alternative optical simulation framework to support the development of DARWIN, the next-generation rare event search with a LXe TPC detector. After establishing a satisfactory agreement with the conventional (Geant4) framework, we performed several studies of light collection efficiency and related metrics. One of the outcomes of these studies is that the default approach for the next-generation detector design, based on scaling up the current-generation detectors to the 50-ton scale, is expected to give a reasonable average LCE value of 43\%, which can be further augmented by iterative design improvements, such as covering the electrodes with VUV reflective coatings. More drastic design modifications, such as the 4$\pi$ variant, lead to a lower LCE in spite of the increased total active area due to the obstruction by the FSRs. 

Optical simulations alone are not enough to fully guide the detector development, which requires a complete simulation that includes energy deposition, electron drift and diffusion, coincidence with dark noise, electronics and other effects. Incorporating the Chroma engine in the simulation framework is a natural avenue for future work. It may also be noted that any simulation is only as good as its input, and more measurements of optical parameters at VUV wavelengths and in LXe environment for materials relevant to the next-generation detector (Tables~\ref{tab:optical},~\ref{tab:optical_4pi}) are needed. 

Nevertheless, photon tracking is often one of the most time-consuming steps, and the ever-increasing size and complexity of detectors only exacerbates this. The framework used in this work allowed us to conveniently import detailed detector design directly from CAD files and track a large number of photons at a time scale that is about an order of magnitude faster than Geant4. As GPU computing becomes more common, many computing clusters begin to offer large GPU farms for research use, further improving performance for such simulations. 

To conclude, the approach to optical simulations explored in this work is an attractive addition to the modern detector developer's toolset and can be instrumental in guiding the design of future experiments~\cite{darwin2016,kt,wp} that use liquid xenon detectors.

\acknowledgments
This work is supported 
by a Department of Energy Grant No~DE-SC0019261,
by the European Research Council (ERC) under the European Union’s Horizon 2020 research and innovation programme, grant agreement No~742789 (Xenoscope) and No~724320 (ULTIMATE),
by the SNF Grant~200020-188716, 
by the European Union’s Horizon 2020 research and innovation programme under the Marie Skodowska-Curie grant agreement No~860881-HIDDeN,
by the University of Zurich,
by the Bundesministerium für Bildung und Forschung (BMBF),
by the Max-Planck-Gesellschaft, 
by the Helmholtz Association, 
by the Deutsche Forschungsgemeinschaft (DFG), 
by National Science Foundation grants PHY-2112803 and PHY-2046549,
by CERN/FIS-TEC/0038/2021 and UIDP/FIS/04559/2020 (LIBPhys), funded by national funds through FCT - Fundação para a Ciência e Tecnologia,
by Pazy Foundation  and Israel Science Foundation (ISF).
We gratefully acknowledge the support of Nvidia Corporation with the donation of three Titan Xp GPUs used for the optical simulations.

\section*{Code and data availability statement}
Relevant code and MC data used in this work can be made available upon request to corresponding authors.

\appendix
\section*{Appendix: Issues found in the original Chroma code}

\begin{enumerate}
\item When a photon is incident on a surface in the original Chroma code, a plane of incidence is determined by taking a cross product of the photon's incoming direction vector and the normal to the surface. An obvious special case is normal incidence, when the plane of incidence becomes undefined. Chroma deals with this special case correctly when treating refraction and diffuse reflection. However, for specular reflection, the original code was ignoring this special case, leading to the propagation code aborting the tracking of such photons. We fixed the issue in our local Chroma installation, even though it had no noticeable impact on LCE due to a low probability of occurrence in a realistic simulation.

\item If two reflecting planes are joined at an angle, then photons that hit the intersection between the planes are not getting reflected. It appears the issue is due to conflicting definitions of the normal to the reflecting surface in such a case. In the TPC simulation this may happen, for example, if a photon hits a surface between two of the 24 segments of the PTFE reflecting panels. This issue has not been fixed. However, it is inconsequential in any realistic simulation, due to the very low probability of such occurrence. It was checked explicitly that replacing the segmented sidewall geometry by a cylindrical one leads to no perceptible difference in average LCE.

\item The Rayleigh scattering algorithm in Chroma was producing a distribution of angles between the scattered and initial photon's direction and polarization vectors that was distinctly different from the one produced by Geant4. Figure~\ref{fig:ray_comparison} demonstrates the difference. Since the Geant4 algorithm (implemented in G4OpRayleigh.cc file) reproduces the theoretical distribution, it has been included in our local copy of Chroma and is now used by default.
\begin{figure}[htpb]
    \centering
    \includegraphics[width=0.49\textwidth]{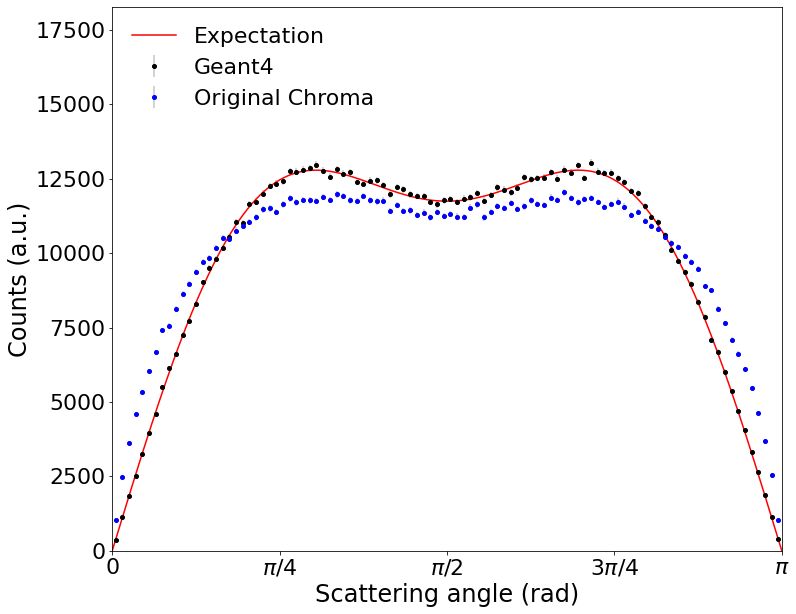}
    \caption{Distribution of angles between the Rayleigh-scattered and initial photon's direction vectors. The original Chroma's algorithm is shown in blue. Geant4's algorithm is shown in black. The red line shows the theoretical distribution. We have fixed the algorithm in our local copy of Chroma to reproduce the Geant4's one.}
    \label{fig:ray_comparison}
\end{figure}
The impact of this issue translated into a couple percent decrease of the average LCE in the Chroma simulation.
\end{enumerate}

\bibliographystyle{JHEP}
\bibliography{main}

\end{document}